\documentclass[aps,prb,epsf,twocolumn]{revtex4}
\usepackage{amsmath}
\usepackage{graphicx}
\usepackage{amssymb}

\newcommand{\beq}{\begin{equation}}
\newcommand{\eeq}{\end{equation}}

\newcommand{\bea}{\begin{eqnarray}}
\newcommand{\eea}{\end{eqnarray}}

\def\tit#1#2#3#4#5{{#1}{\bf #2}, #3 (#4)}

\def\npb{Nucl.\ Phys.\ B\ }

\def\prl{Phys.\ Rev.\ Lett.\ }

\def\jpa{J.\ Phys.\ A\ }

\begin{document}

\title{Resonating singlet valence plaquettes}
\author{S. Pankov}
\affiliation{National High Magnetic Field Laboratory, Florida
State University, Tallahassee, FL 32306, USA}
\author{R. Moessner}
\affiliation{Rudolf Peierls Centre for Theoretical Physics, 
Oxford University, Oxford OX1 3NP, United Kingdom}
\author{S. L. Sondhi}
\affiliation{Department of Physics and Princeton Center for 
Theoretical Physics, 
Princeton University,
Princeton, New Jersey 08544, USA}
\date{\today{}}

\begin{abstract}
We consider the simplest generalizations of the valence bond physics
of $SU(2)$ singlets to $SU(N)$ singlets that comprise objects with $N$ 
sites---these are $SU(N)$ singlet plaquettes with $N=3$ and $N=4$ in 
three spatial dimensions. Specifically, we search for a quantum mechanical
liquid of such objects---a resonating singlet valence plaquette phase
that generalizes the celebrated resonating valence bond phase for
$SU(2)$ spins. We extend the Rokhsar-Kivelson construction of the quantum 
dimer model to the simplest $SU(4)$ model for valence plaquette dynamics on
a cubic lattice. The phase diagram of the resulting quantum plaquette model is 
analyzed both analytically and numerically. We find that the
ground state is solid everywhere, including at the Rokhsar-Kivelson 
point where the
ground state is an equal amplitude sum. By contrast, the equal
amplitude sum of SU(3) singlet triangular plaquettes on the face
centered cubic lattice is liquid and thus a candidate for describing a
resonating single valence plaquette phase, given a suitably defined
local Hamiltonian.
\end{abstract}
\maketitle

\section{Introduction}
The physics of valence bonds \cite{andersonfazekas,anderson87} has
provided a fruitful framework for understanding non-N{\'e}el phases of
$SU(2)$ invariant quantum magnets in low dimensions.\cite{sachrev,misrev}
In particular,
its reformulation as the study of quantum dimer models by Rokhsar and
Kivelson\cite{Rokhsar88} has yielded a sharp but intuitive description
of a diverse set of unconventional phenomena, most notably the
existence of topologically ordered, fractionalised RVB 
liquids\cite{moessner01,gregdidvin}
originally proposed by Anderson and the phenomenon of
Cantor deconfinement\cite{fhmos}.  Extensions of such dimer models to
higher dimensions have been
considered\cite{HuseKrauthMoeSon,HermeleFisherBalents,Henley,SenthilMotrunich,CenkeGrav,JoelCM}
and in the process the existence of two distinct RVB liquids in $d=3$
has been demonstrated. The dimer model formulation has the additional
virtue of having the manifest interpretation of a strongly coupled,
frustrated, gauge theory with RVB phases arising as deconfined phases
in such gauge theories.\cite{fradkiv,moessnersondhifradkin} It is also useful
to note the recent construction of finite range spin models that
realize dimer models arbitrarily accurately in their low energy
sectors.\cite{RamanMoessnerSondhi,FujiKlein} We 
should note that the interest
in RVB liquids also comes from  a much larger study of topological 
phases in condensed matter systems.\cite{xgwbook}

In this paper we pursue a generalization of the valence bond idea to
$SU(N)$ groups with $N >2$ with the aim of searching for topological
phases in models with these higher symmetries. The valence bond is
replaced by a singlet formed by $N$ spins in the fundamental
representation of $SU(N)$. The basic 
geometrical degrees of freedom thus become triangles for $N=3$, and 
objects containing four sites, such as square plaquettes for $N=4$,
and so on.
We restrict our study to singlet plaquettes with $N=3,4$. 
With the assumption
that for suitable microscopic Hamiltonians the low energy sector is 
well described by coverings of the lattice by such objects, we are
led to study quantum plaquette models.  

These plaquette models are similar in spirit to two-form gauge theories, 
where degrees of freedom are also defined on plaquettes.\cite{savitreview}
 However,
there is one important difference. In the latter, there is a local
constraint on plaquettes that share a bond. In our problem, the site
origin of the plaquette variables is reflected in their satisfying
a site constraint---specifically, that one and only one of the plaquettes
that share a common site be occupied by a singlet.

Our strategy in this paper, mirroring that used in the study of valence 
bond physics, is to begin with a Hilbert space spanned by orthogonal
states $|\tilde \alpha \rangle$ labelled by admissible plaquette coverings 
$\alpha$ of 
the lattice under study. In this space we study the phase diagram
of the simplest local Hamiltonian which includes the quantum dynamics of
resonance between pairs of plaquettes and a potential energy of interaction
between them which we shall term the quantum plaquette model (QPM).
As is well known, when the signs of the matrix elements are
of the ``Perron-Frobenius'' form (i.e.\ no positive off-diagonal 
matrix elements), 
such models exhibit a Rokhsar-Kivelson point (which occurs at 
$v=2t$ for the case of
the cubic lattice, Eq.~\ref{RKhamiltonian})
where the equal amplitude superposition,
\beq
|\psi \rangle = \sum_\alpha |\tilde \alpha \rangle
\eeq
saturates a lower bound on the ground state energy of the problem. The
diagonal correlations in this state define a classical statistical mechanics
of plaquettes which can be studied to establish the true nature of this
potentially liquid state. Knowledge of this special point, along with
the analysis of other limits typically allows the phase diagram to be
established with some confidence. The remaining challenge is to show
that one can actually define a sign convention for wavefunctions of
the spins that lead to the desired signs of the matrix elements and finally
to construct local spin Hamiltonians that encode all of this physics.

For the $N=4$ problem on the cubic lattice we report considerable progress 
on this strategy.
We show that matrix elements of the nearest neighbor spin interaction
have the desired form. We analyze the statistics of the RK point
wavefunction computationally, using a plaquette version of the pocket
Monte Carlo algorithm developed for dimers in
Ref.~\onlinecite{krauth03}. We find that the classical system exhibits
long range order, in which the preferable plaquette locations form a
cubic lattice, analogous to the plaquette valence bond crystal for
dimers. This crystalline phase is accounted for analytically from
several perspectives. Besides this plaquette phase, the phase diagram
of the QPM also contains the staggered and columnar phases and thus
the model does not allow a topological phase.
By contrast, we find that the equal amplitude sum for the $N=3$ problem on
the face centered cubic (fcc) lattice is liquid and thus a candidate
for an RSVP phase. The caveat here arises from the non-ergodic nature
of the dynamics in the simplest QPM on this lattice. While we suspect that
a longer ranged (but still local) 
dynamics will fix this, it remains an open problem at
this time.

We note that we are, by no means, the first to consider such higher symmetry
problems and the emergence of plaquette variables---the new element in 
our work is the study of QPMs and the idea of the plaquette liquid.
In $d=1$,
SU(4) symmetric models are readily constructed by fine-tuning an
SU(2)$\times$SU(2) spin-orbital models,\cite{li98,yamashita98,bossche86} 
or a two-leg
ladder.\cite{momoi03} In two dimensions, singlet plaquettes have already
been considered on a triangular lattice.\cite{li98,penc03} Separately,
SU(4) models have been studied {\em en route} to connecting large-$N$
SU($N$) models with the SU($2$) case.\cite{assaad,paramekanti07} 
The prospect of
studying spin-$3/2$ cold atom systems has inspired an elegant body of work that
includes relevant results on SU(4) plaquette solids and even some
generalizations to SU(N) ladders \cite{cwu06,cwu05}. 
It is also useful
to remark here that models with interspersed conjugate representations
on bipartite lattices \cite{sun-readsach} exhibit singlet dimers 
(``mesons'') while we consider models with uniform choices of representation 
which are thus forced to exhibit larger objects (``hadrons'').

This paper is organized as follows. In Sect.~\ref{model} we repeat
Rokhsar and Kivelson's original ``derivation'' of the $N=4$ 
QPM---essentially, it shows how the appropriate signs can be generated
for the matrix elements of the QPM. In
Sect.~\ref{pocket} we briefly introduce the pocket algorithm and
present the results of numerical simulation supporting our view of the
ground state at the RK point of the model.
In Sect.~\ref{meanfield} we provide an analytical
explanation of the plaquette phase based on a low- and
high-dimensional generalisations of the model. In
Sect.~\ref{fccsect}, we present an analysis of the $N=3$ QPM on
the face-centred cubic lattice.  In the conclusion, we comment on
the model's utility and open questions.

\begin{figure}
\includegraphics[width=.9 \columnwidth]{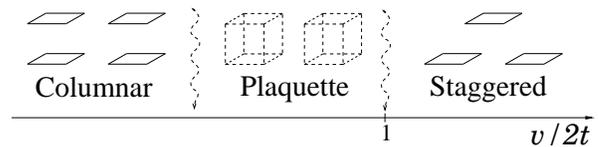}
\caption{Minimal phase diagram of the singlet valence plaquette model
on a cubic lattice. The transitions are first order.}
\label{phase_diag}
\end{figure}

\section{The N=4 QPM on the cubic lattice}
\subsection{Microscopics and Hamiltonian}
\label{model}

The quantum plaquette model is an effective model for
the SU(4) Heisenberg antiferromagnet in a valence plaquette dominated
phase, in much the same way the quantum dimer model is an effective model 
for the SU(2) Heisenberg antiferromagnet in a valence bond dominated
phase. As we have noted above, the choice of plaquettes, rather than
dimers, as new degrees of freedom is dictated by the fact that in the
SU(4) model the lowest energy configuration on a four site lattice is
a singlet which mixes equally all four sites. 

In this section we will establish some useful properties of the
microscopic wavefunctions that correspond to plaquette coverings 
of the cubic lattice. Should it prove possible to construct a
Hamiltonian in the full SU($N$) invariant Hilbert space which has the 
desired coverings, and no other states, as ground states---a non-trivial 
problem---the results in this section can be used to imitate Refs.
\onlinecite{RamanMoessnerSondhi,FujiKlein} and rigorously realize
the QPM that we introduce here. We will not have much to say about this
possibility beyond a few closing remarks in the last section.

A classical plaquette
covering of the lattice corresponds to a quantum state, the wave
function $|\alpha\rangle$ of which is the product of individual
plaquette singlet wave functions. These states are not orthogonal and
computing the Hamiltonian matrix elements requires knowledge of the
overlap matrix elements. In the case of the dimer model, a simple
expression for the overlap matrix elements was given in terms of the
length of loops in the transition graph (which is formed by
superposing the two dimer coverings under
consideration).\cite{sutherland88}

\begin{figure}
\includegraphics[width=.9 \columnwidth]{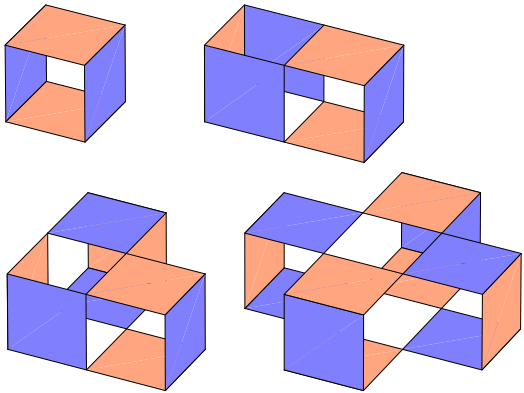}
\caption{(color online) 
Transition graphs with various number of strongly connected
  parts.}
\label{transition_graphs}
\end{figure}

For the plaquette model, the situation is more involved as the
geometry of the transition graph elements is much more complex than
the loops formed by the dimers. We have depicted the simplest members
in Fig.~\ref{transition_graphs}, and it is easy to check
that their contribution to the overlap is given by a factor
\begin{equation}
S=6^{-\frac{l}{2}+1}
\label{overlap}
\end{equation}
Here $l$ is the total number of plaquettes in the transition graph
element. 
With this definition of $l$, the overlap in the SU(2) dimer
case is given by $2^{-\frac{l}{2}+1}$, which shows that the value of
the convergence parameter, $x=1/6$, is much smaller in the SU(4) case
than it is for SU(2), where it equals $1/2$.

In general, configurations contributing to the overlap can be mapped
bijectively onto an edge four-coloring model on the graph in which the
centres of the plaquettes define the sites and their overlapping
corners the edges. For the case of a transition graph element defined
by two overlapping checkerboards, one thus finds that the overlap
matrix element is even lower, as that entropy\cite{JacobsenKondevNPB}
implies $.116^{l/2}$. Interestingly, the Pauling estimate for this
quantity is the same as for the ice model on the same graph, namely
$0.094^{l/2}$, although it is obvious that the entropy of the ice
model is lower. 

We are not aware of a good rigorous bound for the entropy of such loop
models; it is straightforward to show
(Appendix~\ref{overlap_matrix_element}) that $S\le
c\,\left({3^{\frac{1}{3}}}/{2}\right)^{\frac{l}{2}}=0.711^\frac{l}{2}$,
(where $c$ is an unimportant constant) is such a bound, although not
at all a strict one. Indeed, all the cases we have explicitly checked
have Eq.~\ref{overlap} as an upper bound: $S\leq6^{-\frac{l}{2}+1}$.

\begin{figure}
\includegraphics[width=.9 \columnwidth]{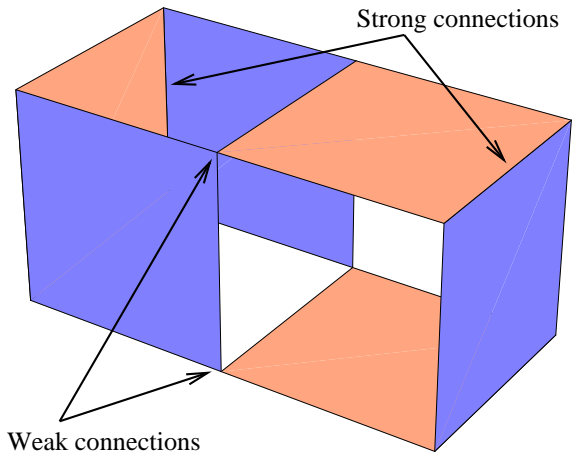}
\caption{(color online) 
Transition graph generated by two configurations (denoted
light and dark, respectively) of three plaquettes. A strong connection
involves two plaquettes sharing an edge, whereas a weak connection is
formed by two plaquettes coinciding at a single site only. }
\label{graph}
\end{figure}

It is convenient to work in the orthogonal basis
$|\tilde\alpha\rangle$ which can be constructed from the original
basis $|\alpha\rangle$ as
$|\tilde\alpha\rangle=(S^{-\frac{1}{2}})_{\alpha\alpha'}|\alpha'\rangle$,
where $S_{\alpha\alpha'}=\langle\alpha|\alpha'\rangle$. Such an
orthogonalization can be conveniently carried out analytically assuming
the off-diagonal overlap matrix elements to be 
small. 

For example, consider the projection of the nearest neighbor
$SU(4)$ Hamiltonian
\beq
H= \sum_{\langle i j \rangle}\sum_{mn}S^n_m(i)S^m_n(j),
\eeq
where the $S^n_m$ are $SU(4)$ generators, to the valence plaquette
subspace.
To this end one expands $S_{\alpha\alpha'}$ or
$H_{\tilde\alpha\tilde\alpha'}=\langle\tilde\alpha|H|\tilde\alpha'\rangle$
in powers of $x$. Retaining the leading off-diagonal (kinetic) and
diagonal (potential) terms, one obtains the QPM 
Hamiltonian (also referred to as the RK model):
\begin{equation}
H=-t\sum_{\tilde\alpha\ne\tilde\beta}|\tilde\alpha\rangle\langle\tilde\beta|+
v\sum_{\tilde\alpha}|\tilde\alpha\rangle\langle\tilde\alpha|
\label{RKhamiltonian}
\end{equation}
where $t=5 x$ and $v=16 x^2$
(see Appendix \ref{RKmodel} for details).

The precise numerical values for $t$ and $v$
should not be taken seriously -- due to the presence of
higher-order corrections and the absence of other simultaneous perturbations
that one can consider, and the choice of plaquette coverings itself
remains to be energetically forced. Nonetheless the negative sign of the
kinetic term suggests that it is reasonable to study 
a QPM with the above ``Perron-Frobenius'' form which is what we
will do next. For such models one has the highly useful feature that 
the ground state is a nodeless wavefunction.

We note that the states $\tilde\alpha$ and $\tilde\beta$ are connected by 
a single flip of a pair of plaquettes. It is easy to show that one
ground state at the RK point ($v=2t$)
is the equal amplitude superposition of all possible states
$|\tilde\alpha\rangle$. Studying its properties reduces to sampling (with
equal probability) different plaquette coverings.
We use a pocket algorithm\cite{krauth03} to implement this task.

\subsection{Pocket algorithm and MC simulation results}
\label{pocket}

The main advantage of the pocket algorithm is that it efficiently
finds possible cluster moves.  The cluster moves, starting from a
covering $A$, are constructed as follows. 1) Consider a transition
graph, {$\cal C$}, between $A$ and another covering, $B$. 2) Pick one
connected part of the transition graph, {$\cal P\in C$}; {$\cal P$} is
the content of the pocket. 3) Replace the plaquettes of $A$ belonging
to ${\cal P}$ by those of $B$. The new covering $A'$ obtained in this
way from $A$ is an allowed covering, because it automatically
satisfies the close packing constraint.

Finding an uncorrelated new covering $B$, is of course in general not
an easy problem. In the pocket algorithm, it is obtained by acting on
$A$ with a symmetry operation\cite{dresskrauth} of the lattice
(reflection, for example). Replacing $A$ with a reflected version of
itself would of course not yield an independent configuration, but
partial replacements (using ${\cal P}$ only) eventually do.

This algorithm has been applied to the dimer
problem.\cite{krauth03} In that case the pocket contains a one
dimensional object which was shown not to invade space, as loops in
dimer transition graphs never branch. The situation is different with
plaquettes. It turns out that the transition graph actually percolates
and the system is not very close to the percolation threshold. In
other words the typical size of non-percolating pocket is small. This,
however, does not pose a problem, as the non-percolating pockets contain 
about $40\%$ of all plaquettes, 
that is only about $60\%$ of the plaquettes are
flipped when the percolating pocket is encountered. 

The limiting factor in our case is the quick growth of the
equilibration time $\tau$ with increasing size of the lattice. The
growth appears to be faster than $L^3$ with $L$ being the size of the
lattice. To establish $\tau$ we used a binning procedure (see Appendix
\ref{binning} for details).  Due to these ergodicity restrictions we
were limited to a linear size $L=64$, i.e.\ a total of $64^3/4=65536$
plaquettes, which is sufficient to establish the ordering properties
of the model.

In the simulations we measure various (equal time) correlation
functions, as a function of plaquette separation. The coordinates
${\bf r}=(x,y,z)$ are assigned to a plaquette covering sites
$(i,j,k,l)$ according to the rule: $x=\min{(x_i,x_j,x_k,x_l)}$,
$y=\min{(y_i,y_j,y_k,y_l)}$ and $z=\min{(z_i,z_j,z_k,z_l)}$, where
$(x_i,y_i,z_i)$ are the coordinates of the site $i$ etc. In the same
way we can assign coordinates to a flippable pair of plaquettes (which
we also call a resonating cube). We use letters $X$, $Y$ and $Z$ to
denote the orientation of the plaquette or the pair forming a
`resonating cube'. A letter corresponds to the axis orthogonal to a
given plaquette (or plaquettes, in case of a resonating cube).  Most
generally we are interested in studying the correlation function
$G_{n\alpha\beta}({\bf r})$, where $n=1$ for the plaquette correlator,
$n=2$ for the resonating cube correlator, and $\alpha$ and $\beta$
denote orientations. In fact we will only need to consider the case of
$\alpha=\beta$. Using cubic symmetry we drop the orientation index and
denote the correlation function as $G_{n}(r_{\perp},r_{\parallel})$,
where $r_{\perp}$ and $r_{\parallel}$ refer to the components of ${\bf
r}$ orthogonal and parallel to the plaquette.

\begin{figure}
\includegraphics[width=.9 \columnwidth]{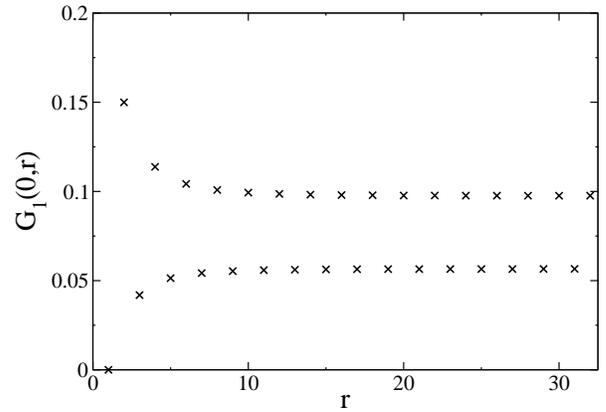}
\vspace{.3cm}
\caption{Correlation function $G_1(0,r)$ for $L=64$.}
\label{c0rl64}
\end{figure}

\begin{figure}
\vspace{.5cm}
\includegraphics[width=.9 \columnwidth]{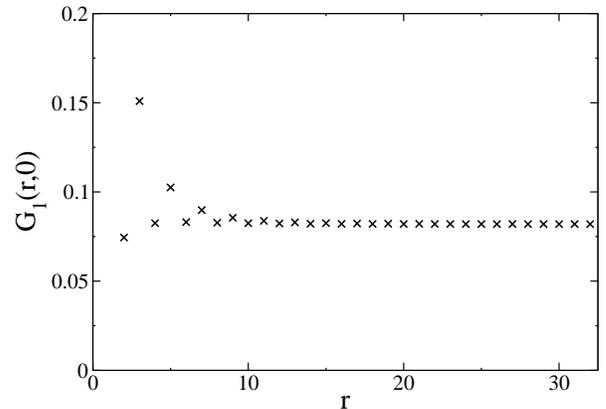}
\vspace{.3cm}
\caption{Correlation function $G_1(r,0)$ for $L=64$.}
\label{cr0l64}
\end{figure}

In Fig.~\ref{c0rl64}, we plot the plaquette correlation function
$G_1(0,r)$. It displays oscillations persisting to large separation
$r$, so that one may conclude that the system possesses long range
order. The simplest possibility might be some kind of
orientational order (rotational symmetry breaking).
Quite a different conclusion may be drawn from the plot
of $G_1(r,0)$, shown in the Fig.~\ref{cr0l64}. At face value, the plot
suggests an absence of the long range order, because the oscillations
of the correlation functions decrease fast with distance. To detect a
possible presence of orientational order we have computed the quantity
$\gamma=|{\bf n}{\bf e}^{(3)}|$, where the components of ${\bf n}$ are
portions of plaquettes of various orientations and the components of
${\bf e}^{(3)}$ are cubic roots of unity. That is ${\bf
n}=(n_X,n_Y,n_Z)$, $n_X+n_Y+n_Z=1$ and ${\bf
e}^{(3)}=(1,e^{i2\pi/3},e^{i4\pi/3})$. In the case of orientational
order, $\langle\gamma\rangle$ would approach a finite value as size of
the system increases. In our case it scales to $0$ instead.

\begin{figure}
\vspace{.3cm}
\includegraphics[width=.9 \columnwidth]{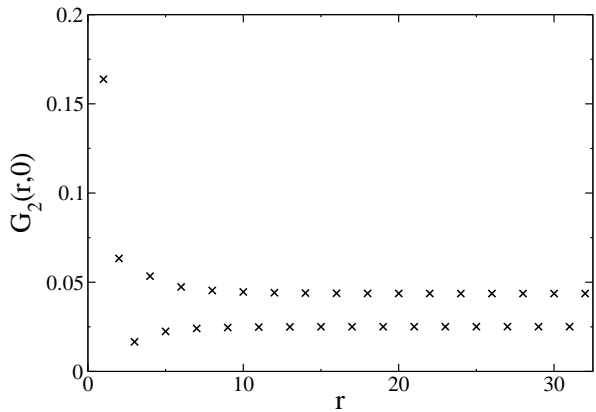}
\vspace{.3cm}
\caption{Pair correlation function $G_2(r,0)$ for $L=64$.}
\label{pcr0l64}
\end{figure}

\begin{figure}
\includegraphics[width=.9 \columnwidth]{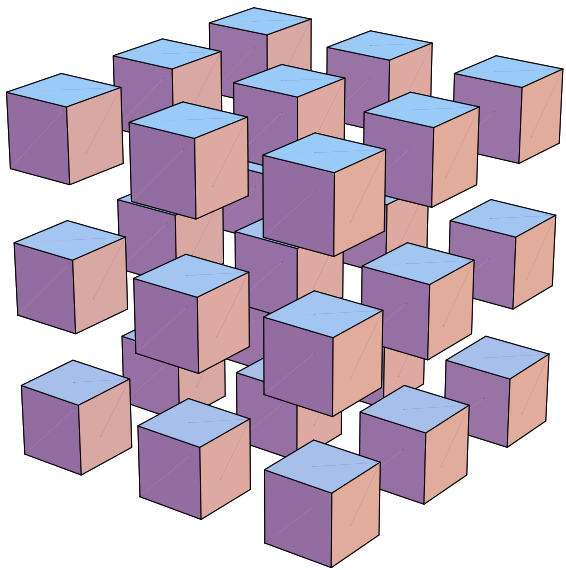}
\vspace{.3cm}
\caption{(color online) 
Ordering pattern. Plaquettes form resonating cubes which are
ordered on a cubic lattice with doubled lattice spacing. Different
colors represent plaquettes of three different orientations. The
plaquette orientations are uncorrelated over large distances.}
\label{order_pattern}
\end{figure}

More insight can be gained from the resonating cube correlation
functions. In Fig.~\ref{pcr0l64} we plot the function
$G_2(r,0)$. Unlike the function $G_1(r,0)$, it clearly captures strong
persistent correlations in the direction perpendicular to the
plaquette as well. Our findings for $G_1$, $G_2$ and $\gamma$ can be
combined in a coherent picture in which the resonating cubes
form a lattice, breaking translational symmetry, but the orientations
of plaquettes on different cubes are not correlated over long
distances -- thus orientational symmetry is not broken. This is
illustrated in Fig.~\ref{order_pattern}.

\begin{figure}
\vspace{.5cm}
\includegraphics[width=.9 \columnwidth]{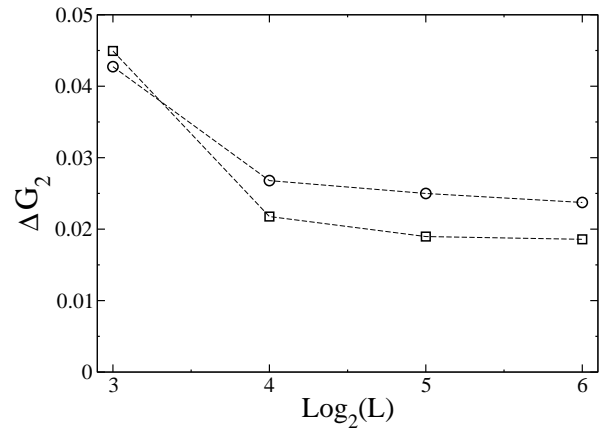}
\vspace{.3cm}
\caption{The curves show scaling of $G_2$ with the size. $\Delta
G_2(0,L/2)$ (circles) and $\Delta G_2(L/2,0)$ (squares)
are plotted vs $\log_2{(L)}$. Lines are guides to the eye.}
\label{scalingG2}
\end{figure}

The presence of long range order
(LRO) suggested by Fig.~\ref{pcr0l64} 
is supported by  a scaling analysis. In 
Fig.~\ref{scalingG2} we plot differences $\Delta
G_2(0,L/2)=G_2(0,L/2+1)-G_2(0,L/2)$ and $\Delta
G_2(L/2,0)=G_2(L/2+1,0)-G_2(L/2,0)$ for $L=2^p$. 
We see that $\Delta G_2$
approaches a constant already at $p\sim 5-6$. [For larger $p$, 
we lose
ergodicity, as we show in Appendix~\ref{binning}, where we evaluate
the relevant equilibration time $\tau$ using a binning procedure.]

\subsection{Origin of the ordered pattern}
\label{meanfield}

Our MC simulations revealed a crystalline phase in the classical
system of plaquette coverings of 3D square lattice. The system
orders in a cubic lattice pattern with the periodicity equal to
two lattice constants of the original lattice.

\begin{figure}
\includegraphics[width=.9 \columnwidth]{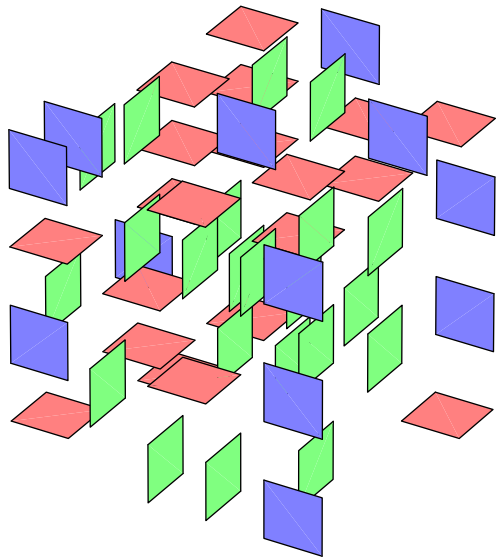}
\vspace{.0cm}
\caption{(color online) 
Snapshot of a plaquette configuration on a cubic lattice}
\label{cubicsnapshot}
\end{figure}

This ordering is tenuous, as evidenced by the small value of the
order parameter. In fact, when one judges a snapshot of the Monte
Carlo simulations by eye (Fig.~\ref{cubicsnapshot}), 
no ordering is evident.

One way to understand the appearance of the long range order is to
consider a more general model in which $n_d$-dimensional hypercubes
are arranged on a $n_s$ dimensional lattice. For RSVP, $n_d=n_s=3$.

Let us start with $n_s=1$, hypercubes arranged on a chain,
Fig.~\ref{chain_fig}. One can then explicitly compute (see details in
the Appendix \ref{chain}) the correlation length, which grows as
$\sqrt{n_d}$ with $n_d$. This result can be visualised by noting that
a resonating cube can be broken up into two domain walls by fixing the
plaquettes to be perpendicular to the chain direction. This costs an
entropy $S_d=\ln n_d$. The domain walls can then be separated at will
without disturbing other cubes. In a system of length $L$, this leads
to an entropy of $S_w=\ln ({L\atop2})$. The correlation length is then
determined by the smallest $L$ at which the introduction of domain
walls becomes favourable ($S_w>S_d$), which happens at
$L=\xi\sim\sqrt{n_d}$.

While there is no true long range order in one dimension, this result
shows that the ordering tendency increases with $n_d$. In this spirit,
the case $n_d=n_s=2$, square dimers, being critical, is indeed already
closer to ordering. Zeng and Henley, for the honeycomb lattice, have
developed a prescription very similar to the introduction of a limit
of large $n_d$ for $n_s=2$, and they found numerically the existence
of an ordering transition for sufficiently large $n_d$.\cite{zeng97}

\begin{figure}
\includegraphics[width=.9 \columnwidth]{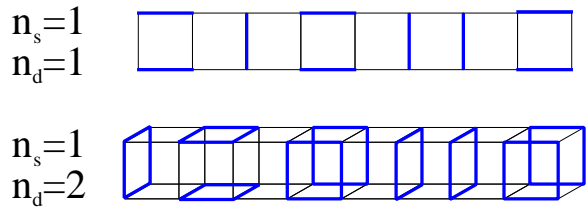}
\vspace{.3cm}
\caption{(color online)
$n_d$ dimensional `hypercubes' resonating on a $n_s=1$ dimensional
chain.
}
\label{chain_fig}
\end{figure}

Thence, whereas for $n_s=1$, $n_d\rightarrow \infty$ yields a critical
point, $n_s=2$ requires only a finite $n_d$ to produce ordering. Our
results adds the insight that for $n_s=3$, $n_d=n_s$ is already
sufficient to  produce ordering.

Some further understanding can be gained by considering higher spatial
dimensionality as chains of $d=n_d$-dimensional hypercubes,
transversally coupled in $d-1$ transverse directions.  If a chain and
its neighbour (in the $\hat{e}_i$ direction, say) are in registry, a
pair of neighbouring resonating cubes can gain further free energy
when their plaquettes are perpendicular to $\hat{e}_i$. This happens
in 1 out of $d$ configurations for each cube, and the number of
additional states of the two neighbouring plaquettes belonging to
different chains resonating together is $d$. This thus yields an
attractive interaction of strength proportional to $1/d$.

For a $d$ dimensional interaction of strength $J\times 1/d$, the
critical coupling $J$ would be $O(1)$.  Our interaction, however, is
amplified by the length of the domains, $\xi\sim\sqrt{d}$, over
which individual chains are ordered even in the absence of transverse
coupling. This means that for large $d$, the transversally coupled
chains will order. This establishes at least local stability of the
plaquette ordering for large $d$, to which our case $d=3$ (but not
yet the case $d=2$) is connected.

Note that this ordering pattern does not imply that, even at large
$d$, the plaquettes spend almost all their time resonating with their
partner on one given cube. (Such cubes define the 'site' of the
ordering lattice depicted in Fig.~\ref{order_pattern}). Rather, they
spend half of their time resonating on the $d$ bonds emanating from
sites of that lattice, as can be seen from an appropriate adaptation
of the above argument.

Let $p_0\ (p_1)$ be the respective probabilities of finding a crystal
site (bond) to be occupied by two parallel plaquettes. The probability
of a resonating cube to appear on any given bond is proportional to
$p_0^2/d^2$. It should be equal to the probability for a resonating
cube to leave a bond, which is proportional to $p_1/d$. In the large
$d$ limit other processes are of higher order in $1/d$. Indeed, one
can evaluate probability $p_i$ for a resonating cube to appear in
locations which are $i$ manhattan steps away from crystal sites
(manhattan distance here is measured in original lattice spacings, so
a bond is one manhattan step away from crystal sites). The relation
between $p_i$ and $p_{i+1}$ is similar to the relation between $p_0$
and $p_1$, that is $p_i^2/d^2\sim p_{i+1}/d$ (as long as $i\ll
d$). Therefore $p_i\sim 1/d^{2^i-1}$, while for the number $n_i$ of
locations at manhattan distance $i$ we have $n_i\propto d^i$. It
follows that only terms with $i=0,1$ will contribute in $d\to\infty$
limit, so a simple relationship holds: $p_1 d=(1-p_0)/2$. From there
we find: $p_0=1/2$ and $p_1=1/(4d)$.

\begin{figure}
\includegraphics[width=.9 \columnwidth]{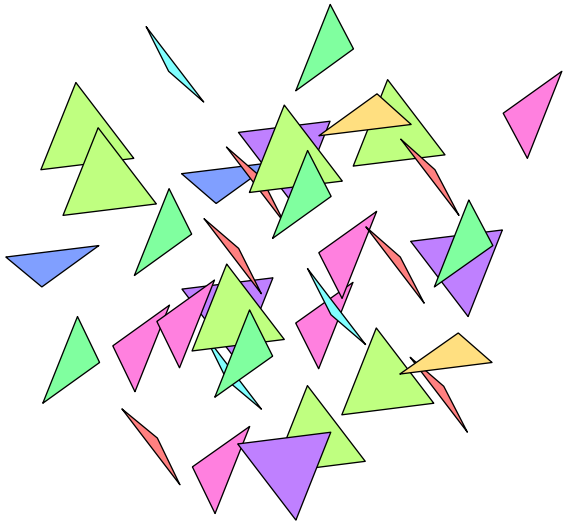}
\vspace{.0cm}
\caption{(color online) 
Snapshot of a palquette configuration on a FCC lattice. 
Eight colors are used to 
represent the different plaquettes, 
as defined by faces of an octahedron (see Fig.\ref{octahedron}).}
\label{fccsnapshot}
\end{figure}

Because a plaquette ia as likely to be found resonating on a bond as
it is on a cube, a random snapshot of the cubic lattice will not
appear to be ordered, similarly to a snapshot of the face centered
cubic (FCC) lattice (Figs.\ref{cubicsnapshot},\ref{fccsnapshot}
and Sect.~\ref{fccsect}). A proper analysis reveals, however, that
plaquette correlations on the FCC lattice are short range, so the
presence of ordering is not a feature of resonating plaquettes in
general, but should be attributed to the lack of frustration on the
cubic lattice.

\subsection{Phase diagram of the RK model}

The above has established that the equal amplitude sum, which 
has zero energy and thus saturates a lower bound on the ground state energy 
at the RK point, is crystalline. The QPM Hamiltonian is not ergodic
in the full Hilbert space---it moves in topological sectors which we
will discuss elsewhere \cite{workinprogress}---and thus will properly
exhibit multiple ground states. We can conclude from the simulation
of the classical ensemble though, that none of these ground states will 
be liquid.

The remainder of the phase diagram of the RK model is completed in the
standard way. For $v/t>2$, any configuration without resonating cubes
becomes a ground state. The ordered plaquette phase takes over in a
first-order transition at the RK point $v/t=2$. This is unlike the
dimer model, in which there is a moderately exotic deconfined multicritical
transition on the square lattice into a solid and into a Coulomb phase
on the cubic
lattice.\cite{moessner01,moessnersondhifradkin,fhmos}

For $-v/t\gg 1$, one finds the columnar solid phase, which is unique
(up to global symmetry operations) in this case. Whether there are
intervening phases between columnar and plaquette solid has not been
studied.

\section{N=3 QPM on the face-centred cubic lattice}
\label{fccsect}

\begin{figure}
\includegraphics[width=.9 \columnwidth]{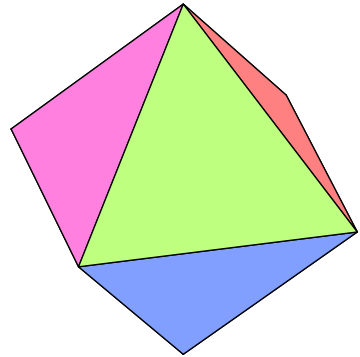}
\vspace{.0cm}
\caption{(color online) 
Octahedron defining the plaquette types on the FCC lattice.}
\label{octahedron}
\end{figure}

To see if the absence of disordered phases is a general feature of
such plaquette models, we have also investigated the case of the
face-centred cubic lattice. There, the elementary plaquettes are
equilateral triangles, the corners of which are defined by a triplet
of mutually nearest-neighbour sites.  The QPM now
includes resonance of a pair of plaquettes the six vertices of
which define an octahedron shown in Fig.\ref{octahedron}.

\begin{figure}
\vspace{.6cm}
\includegraphics[width=.9 \columnwidth]{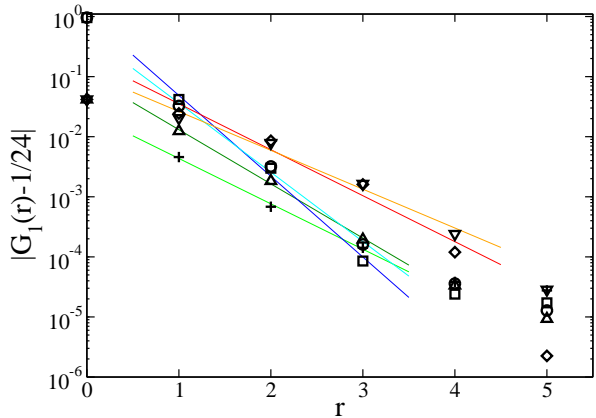}
\vspace{.0cm}
\caption{(color online)
Connected part of the correlation function $G_1(r)$ for various plaquette orientations on the FCC lattice. The distance $r$ is measured in units of the nearest neighbor distance.}
\label{cf_011}
\end{figure}

In Fig.~\ref{cf_011}, we plot the plaquette correlations of the
classical model corresponding to the RK point for various inequivalent
combinations of pair orientations and separations in the [011] direction.
There are six inequivalent correlation functions for four types of 
plaquettes adjacent at an octahedron vertex. All correlators
decay with distance to the value given by uncorrelated
plaquettes. This decay is a rapid exponential, with correlation
lengths of less than a lattice spacing, ranging from $\sim0.3$ (diagonal
component) to $\sim0.7$ (off-diagonal component) 
of nearest neighbor distances.

This demonstrates that a classical hard-core plaquette model is in a
strongly disordered phase and we would be tempted to conclude, by
standard arguments, that we have identified a point in an RSVP phase
that sits at a first order transition to a staggered crystal. 
The connection of the classical
sum to the RK point of the QPM, however, is more complicated than in the 
case of the cubic lattice. The plaquette pair flip is very constrained
and by itself breaks up the Hilbert space into a very large number of
largely frozen states with local quantum fluctuations so that it is 
only the sum of these individual states that exhibits liquidity.
It seems very likely that the inclusion of resonance dynamics on
larger clusters will remove this pathology and establish the RSVP state.
However, at present we are unable to explicitly solve this problem.

\section{Concluding remarks}

Can the above quantum plaquette models be realised in experiment?
Conceivably, they can arise in the case of low energy Ising degrees of
freedom which reside on the midpoints of plaquettes of a lattice, and
which obey the appropriate hardcore constraint. Their simplest quantum
dynamics would then be that of a QPM.

Regarding our original motivation of finding degrees of
freedom with a native SU($N$) symmetry, we first observe that
following initial work on dimer ground state ensembles of Klein
models\cite{ChayesChayesKivelson}, there now exist prescriptions for
local SU(2) invariant Hamiltonians which produce RK quantum dimer
models in a controlled
fashion.\cite{RamanMoessnerSondhi,FujiKlein,kleinzohar}

However, no simple generalisation of Klein models for SU(4) models
with plaquette ground states is known yet, in part for the 
reason that was pointed out in Ref.~\onlinecite{ChayesChayesKivelson}
for dimers on the square lattice. There, the projectors occurring in
the Hamiltonian allow ground states containing not only dimers on
nearest neighbour but also on next nearest neighbour bonds. In an
anlogous manner, for SU(4) plaquette ground states, we are also lead
to other (e.g. non-planar) units of four spins forming a
singlet. Actually realising an SU(4) degree of freedom is of course
also no trivial matter in $d>1$, as combining an orbital pseudospin
1/2 degree of freedom with an SU(2) to form an SU(4) invariant
Hamiltonian requires a fine-tuning of interaction parameters which is
not immediately suggested by the natural symmetries of orbitals
making up the pseudospin 1/2. Similarly, whether or not there exist
lattice models realising an SU(3) QPM on the face-centred
cubic lattice that exhibits an RSVP phase is left for further
investigation.

It is worth noting that SU(4) spins do of course not present the 
exclusive entry point to plaquette degrees of freedom. We have already
referred to work on spin-$3/2$ SU(2) systems that give rise to 
plaquettes \cite{cwu06,cwu05}. One can also consider that
in the same sense
in which dimers can be thought of as low-energy composite degrees of
freedom consisting of a pair of resonating SU(2) spins with $S=1/2$, a
plaquette might be considered as a composite of a pair of resonating
dimers. There is thus nothing fundamentally untoward about resonating
valence plaquettes -- this view being reinforced by the findings of
Ref.~\onlinecite{penc03}, who found plaquette ordering with the same
$\sqrt{12}\times\sqrt{12}$ supercell proposed for resonating dimers on
the same lattice.\cite{MoessnerSondhiIQF,mscPRL00} Similarly,
attractive interactions in the simple cubic dimer model lead to an
ordered phase, albeit one with columnar order,\cite{alet3d} rather
than the plaquette order observed here.

Finally, our work on the RK point also provides an entry to the study
of classical hard plaquettes in $d>2$. In particular, the topological
and ergodicity properties of these models should present an
interesting topic for further research.

\section*{Acknowledgements}

We thank B. Dou{\c c}ot, D. Huse, W. Krauth, F. Mila, M. Mostovoy,
K. Penc and M. Troyer for useful discussions. We are very grateful to
W. Krauth for collaboration on closely related work. This work was
supported in part by the Minist\`ere de la Recherche with an ACI
grant. Most of it was undertaken at Laboratoire de Physique
Th\'eorique de l'Ecole Normale Sup\'erieure, CNRS-UMR 8549.

\appendix

\section{Plaquettes on the chain}
\label{chain}

We use the transfer matrix method to compute correlation function
of $d$ dimensional plaquettes on a chain. 
Each plaquette can
assume $d$ orientational states: one orthogonal to the direction
of the chain and $d-1$ parallel. We
write the Hamiltonian as $H=\sum_i H_{i,i+1}$ and compute the
transfer matrix $T_i^{\alpha\alpha'}= \langle\alpha|\exp{(-\beta
H_{i,i+1})}|\alpha'\rangle$, where $|\alpha\rangle$ represents
local degrees of freedom. Provided the Hamiltonian is
translationally invariant the correlation function $C(x)=\langle
M_iM_{i+x}\rangle$ of some operator $M$ can be written as:
\begin{equation}
C(x)=\frac{\mbox{Tr}M T^x M T^{N-x}}{\mbox{Tr} T^{N}}
\label{cx}
\end{equation}
where $N$ is the length of the chain (periodic boundary conditions
are assumed). In the thermodynamic limit $N\to\infty$ we have:
\begin{equation}
C(x)=\sum_{\alpha}\langle\lambda_0|M|\lambda_{\alpha}\rangle
\langle\lambda_{\alpha}|M|\lambda_0\rangle
\left(\frac{\lambda_{\alpha}}{\lambda_0}\right)^x
\label{cxtherm}
\end{equation}
where $|\lambda_{\alpha}\rangle$ are eigenstates of the transfer
matrix and $\lambda_0$ is its largest (in absolute value) eigenvalue.
The asymptotic form of the connected part of the correlation function
is determined by the next largest eigenvalue $\tilde\lambda$ such
that $\langle\lambda_0|M|\tilde\lambda\rangle
\langle\tilde\lambda|M|\lambda_0\rangle\ne 0$,
so that the correlation length is given by
$\xi=1/\ln{|\lambda_0/\tilde\lambda|}$.
In our case all states are degenerate, so the transfer matrix is
simply proportional to $2d-1$ dimensional matrix of the following
form:

\begin{equation}
T=\left(
\begin{array}{ccc}
1 & A & 0 \\
0 & 0 & E \\
B & C & 0 \\
\end{array}
\right)
\label{tmatrix}
\end{equation}
where $E$ is $(d-1)\times(d-1)$ identity matrix, and all the matrix
elements of $A$, $B$ and $C$ are equal $1$. The matrix $C$ is
$(d-1)\times(d-1)$ square matrix, while $A$ and $B$ are
$1\times(d-1)$ and $(d-1)\times1$ matrices respectively.
It is not hard to find that $T$ has only two nonzero eigenvalues:
\begin{equation}
\lambda=1/2\pm\sqrt{d-3/4}
\end{equation}
from where we find $\tilde\lambda/\lambda_0=-1+1/\sqrt{d}+O(1/d)$.
The first term encodes the fact that the connected part of the
correlation function oscillates, while the second term determines the
correlation length $\xi=\sqrt{d}$.

\section{Overlap matrix element}
\label{overlap_matrix_element}

In the fundamental representation four basis states (flavors) are
used to represent an SU(4) spin. The invariance of the SU(4) singlet
plaquette wave function under rotations in spin space dictates
the singlet to be a fully antisymmetrized combination of all four
flavors on four sites:
\begin{equation}
|\psi_s\rangle=\frac{1}{\sqrt{24}} \sum_{P\{1234\}}(-1)^p
|1234\rangle \label{singletwf}
\end{equation}
where the four flavors $\{1234\}$ are permuted in all possible ways
and $p$ is the parity of permutation $P$. In each term there is one
flavor per site, that is flavors are permuted among different
sites. The prescription for fixing overall sign of the wave
function will be given below. We consider lattice states
which are products of singlet wave functions corresponding to each
plaquette. We need to compute the overlap matrix element between
two such states. It can be visualized with the help of the
transition graph - the graph which is made of all partially
overlapping plaquettes. Only those have to be included, as the
fully overlapping plaquettes multiply the matrix element by $1$.
The total overlap matrix element is a product of overlap matrix
elements corresponding to disconnected parts of the transition
graph, therefore it is sufficient to consider only a connected
transition graph.

Two plaquettes can be connected either via a common edge or a
single common site. We call these strong and weak connections,
respectively. The common edge in the strong connection is called
connecting edge, and the common sites in weak connections are
called connecting sites. If it is possible to connect two
plaquettes by a path belonging to the transition graph and going
through strong connections only, then two plaquettes are said to
belong to the same strongly connected part. Using the fact that
each site in the transition graph is shared by exactly two
plaquettes, it is easy to check that either the strongly connected
part has exactly four weak connections attached, or it has no
weak connections and we call it a simple graph.

\begin{figure}
\includegraphics[width=.9 \columnwidth]{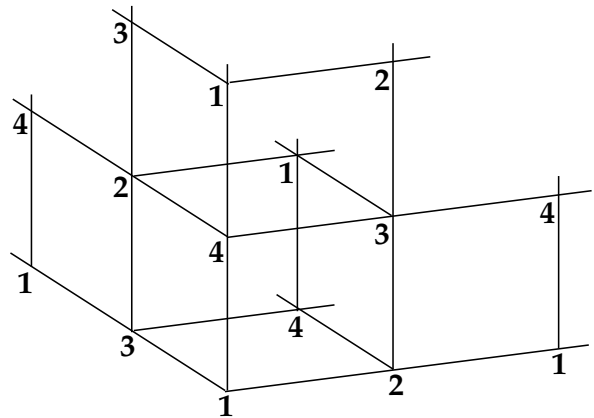}
\caption{Assignment of flavors which respects the constraint of no
two sites in a plaquette carrying same flavor.}
\label{enumeration}
\end{figure}

Each plaquette in the transition graph corresponds to a certain term
$|1234\rangle$ of the singlet wave function. The overlap is non zero
only when sites shared by different plaquettes carry identical flavors
in each of the two plaquettes. Thus a nonzero overlap is parameterized
by assigning a flavor per site in the transition graph, under
the constraint that no two sites in a plaquette carry the same
flavor. This is always possible to do by assigning flavors to all
sites of the lattice in the following way~: assign $1$ to the point
$(0,0,0)$, $2$ to the point $(1,0,0)$, $3$ to the point $(0,1,0)$, $4$
to the point $(0,0,1)$ and then translate this configuration by
$(\pm1,\pm1,\pm1)$ to cover all sites of the lattice (see
Fig.~\ref{enumeration}). It is easy to check that any plaquette on the
lattice contains four different flavors.

It has to be decided how to choose signs for plaquette singlet wave
functions to ensure that the overlap matrix elements of the initial
and final state of a resonance process are positive. It is
straightforward to check that it is enough to require the term
$|1234\rangle$ in each singlet wave function, whose flavors coincide
with our prescription, to be positive.

The assignment of flavors to the sites in the transition graph is not
unique. This is most easily seen by making a connection between this
assignment and a loop model, which in turn establishes the link to the
four-colouring model mentioned in the text. We
partition the sites in the graph (for a given allowed assignment of
flavors) in the following way.  Connect sites with flavors $1$ and $2$
which belong to the same plaquette. Do the same with the sites
carrying flavors $3$ and $4$.  In this way, all sites will be divided
into noncrossing loops.  Notice that in each loop two flavors can be
exchanged without violating the constraint. The configurations of
flavors which can be connected by a sequence of such transformations
are said to belong to the same loop decomposition.\cite{fn-topomoves}
The number of such configurations is $2^{n_l}$, where $n_l$ is the
number of loops in the decomposition. Some loops contain weak
connections, while others do not -- we call them nonlocal and local
loops, respectively. From the flavor constraint (any plaquette includes
all four flavors) it follows that a nonlocal loop goes through at
least two strongly connected parts and has at least length $l=4$,
while a local loop belongs to a single strongly connected part and has
length $l=2$. Obviously, $n_l=n_{ll}+n_{nl}$, where $n_{ll}$ and
$n_{nl}$ is the number of local and nonlocal loops
correspondingly. Taking into consideration that $n_{nl}\le n_s$ and
$n_{ll}^i\le l_i-1$, where $n_{ll}^i$ and $l_i$ are the number of
local loops and the number of plaquettes in the $i$th strongly
connected part (the case $n_s=1$ is not included here), we can write
the following inequality for the contribution $\tilde S$ to the
overlap matrix element from a given loop decomposition:

\begin{equation}
\frac{1}{6}\tilde S=\left(\frac{1}{24}\right)^{\frac{l}{2}} 2^{n_l}
\le 6^{-\frac{l}{2}}
\label{overlap1}
\end{equation}

The prefactor of $1/6$ accounts for various ways of distributing
four flavors among two classes of loops. 
Notice that for a simple graph $n_{ll}=l$ and
$n_{nl}=0<1=n_s$, so the above result still holds in that case,
with the inequality turning into an equality.

In the next step we derive a bound on the number of loop
decompositions. For the sake of compactness we use terms
``vertex" and ``connection" in place of ``strongly connected
part" and ``weak connection" below. Four connections belonging
to the same vertex are joined by nonlocal loops. This can be
done in three different ways. Therefore there are at most
$3^{n_s}$ loop decompositions. Clearly, not all decompositions
are allowed, because for two nonlocal loops going through the
same vertex (there are at most two such loops) a choice of
flavors of one of the loops uniquely determines flavors of
the second loop. This restriction can be formulated as a problem
of coloring loops with just two colors, let us say black and white.
We start with no nonlocal loops defined, that is all connections
disjoined and not colored. We will be joining four connections
(belonging to the same vertex) at a time, and color them according
to the way they are joined. Without loss of generality, we
choose to do the joining procedure on those vertices which
already have colored connections.

In this way colors are always defined uniquely (apart from the
first step). It is easy to check that there are at most
three ways to join connections when none or one of them has been
colored. There are at most two ways of joining connections
when two of them have been colored. There is at most one way
of joining, when three or all connections have been colored.
Thus the upper bound on the number $M$ of loop decomposition
can be written as:

\begin{equation}
M=\max{\left\{2^{n_s}\left(\frac{3}{2}\right)^{n_0+n_1}
\left(\frac{1}{2}\right)^{n_3+n_4}\right\}}
\label{M}
\end{equation}
with the following constraints imposed
\begin{eqnarray}
&& \sum_{i=0}^{i=4}n_i=n_s
\nonumber\\
&& \sum_{i=0}^{i=4}i n_i=2 n_s \label{constraints}
\end{eqnarray}
where $n_i$ is the number of vertices which undergo the joining
procedure having $i$ colored connections, and thus is positive.
From Eq.~\ref{constraints} we have $2 n_0+n_1=n_3+2 n_4$, so the
maximum in Eq.~\ref{M} is achieved when $n_0$ and $n_3$ are
minimized, while $n_1$ and $n_4$ are maximized. In the best case
$n_1/2=n_4=n_s/3$, and $M(n_s)=3^{2n_s/3}$. Since $n_s\le l$, the
bound for $S$ equals the bound for $\tilde S$ times $M(l)$:

\begin{equation}
S\le 6\left(\frac{3^{\frac{1}{3}}}{2}\right)^{\frac{l}{2}}
\label{boundS}
\end{equation}

\section{Matrix Element Computations for SU(4) QPM}
\label{RKmodel}

We use $|\tilde\alpha\rangle$ notations for the orthonormal states
and $|\alpha\rangle$ for the original states. We only need to
consider states which differ by orientation of a single pair of
plaquettes, we thus name the states according to the orientation
of the pair in the YZ, ZX and XY planes as $|X\rangle$, $|Y\rangle$
and $|Z\rangle$. We first compute matrix elements
$\langle\alpha|H|\beta\rangle$. It is easy to
check that:
\begin{equation}
H_{ij}|\alpha(ijkl)\rangle=-|\alpha(ijkl)\rangle
\label{Hijacting}
\end{equation}
where $H_{ij}=\sum_{mn}S^n_m(i)S^m_n(j)$ is the part of the
Hamiltonian acting on the bond $(ij)$. The location of the
plaquette is explicitly shown in $|\alpha(ijkl)\rangle$ and
$S^n_m$ are $SU(4)$ generators.\cite{li98} In the fermionic
representation $S_m^n=a^{\dagger}_m a_n$, so
$S^n_m|\mu\rangle=\delta_{n\mu}|m\rangle$ where $\mu$ is the
flavor. The term $S^n_m(i)S^m_n(j)$ simply exchanges flavors $m$
and $n$ on sites $i$ and $j$ from where the Eq.(\ref{Hijacting})
follows. We note that only $H_{ij}$ with $i$ and $j$ belonging to
any of the two plaquettes need to be considered. When
$\alpha\ne\beta$ there is always a plaquette containing both $i$
and $j$, therefore for any bond
$\langle\alpha|H_{ij}|\beta\rangle=-x$ and
$\langle\alpha|H|\beta\rangle=-12 x$ because twelve bonds belong
to the pair of plaquettes. When $\alpha=\beta$ only eight bonds
belong to an individual plaquette, these bonds give
$\langle\alpha|H_{ij}|\alpha\rangle=-1$. For a bond shared by the
plaquettes one can estimate $\langle\alpha|H_{ij}|\alpha\rangle=
\sum_{\beta\ne\alpha}|\langle\alpha|\beta\rangle|^2
\langle\beta|H_{ij}|\beta\rangle=-2x^2$. So we have
$\langle\alpha|H|\alpha\rangle=-8-8x^2$.

Let us compute, for example, $\langle\tilde X|H|\tilde X\rangle$ and
$\langle\tilde X|H|\tilde Y\rangle$. To the lowest order $|\tilde
X\rangle=|X\rangle-\frac{x}{2}(|Y\rangle+|Z\rangle)$ and $|\tilde
Y\rangle=|Y\rangle-\frac{x}{2}(|X\rangle+|Z\rangle)$.  Then
$\langle\tilde X|H|\tilde X\rangle=\langle
X|H|X\rangle-\frac{x}{2}(\langle X|H|Y\rangle+\langle
X|H|Z\rangle+h.c.)+...=-8+16 x^2+{\cal{O}}(x^3)$. The constant term
$-8$ is actually independent of plaquette configuration and can thus
be dropped. For the off diagonal element we find $\langle\tilde
X|H|\tilde Y\rangle=\langle X|H|Y\rangle-\frac{x}{2}(\langle
X|H|X\rangle+\langle Y|H|Y\rangle)+...=-5 x+{\cal{O}}(x^2)$. Thus we
obtain for the kinetic and potential term in the RK Hamiltonian $t=5x$
and $v=16x^2$.

\section{Determination of the equilibration time}
\label{binning}

\begin{figure}
\includegraphics[width=.9 \columnwidth]{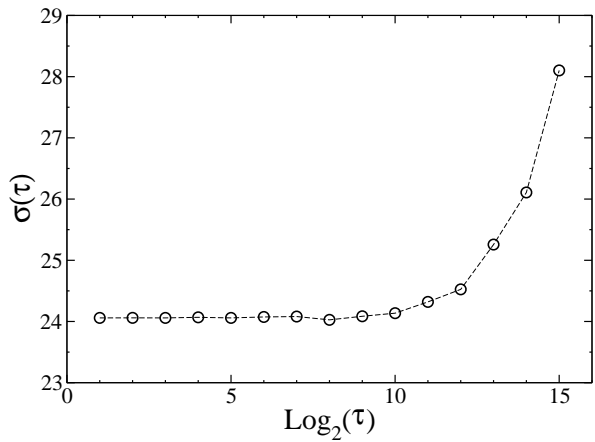}
\caption{The standard deviation $\sigma(\tau)$ of
$\langle G(0,L/2)\rangle_{\tau}$ plotted vs ${\rm Log}_2\tau$ for $L=48$.}
\label{sigma_l48}
\end{figure}

The error of statistically independent measurement decreases with the
number of measurements $n$ as $n^{-\frac{1}{2}}$. If, however, the
measurements are not independent, the error cannot be reduced beyond
certain limit, as more measurements do not add any new
information. This allows to estimate correlation time corresponding to
some physical quantity $O$ by measuring fluctuation of the average
value $\langle O\rangle_{\tau}$ computed over the time span $\tau$.
More precisely, we measure $\langle O\rangle_{\tau}$ over $m$
successive intervals of time of length $\tau$. From $m$ average values
we compute the standard deviation $\sigma(\tau)$ of the distribution
of $\langle O\rangle_{\tau}$. In the same run we can measure $\langle
O\rangle_{\tau/2}$ in $2m$ time intervals by subdividing each $\tau$
interval in two. The corresponding standard deviation $\sigma(\tau/2)$
is thus
computed from $2m$ measurements.  The procedure of subdivision
(of creating smaller bins, hence the term binning) can be continued as
far as needed.  By plotting $\sigma(\tau)$ vs ${\rm Log}_2\tau$ we
estimate the equilibration time $\tau_{eq}$ as the time where
$\sigma(\tau)$ stops changing as $\tau$ is lowered. Alternatively, one could keep the number $m$ of intervals fixed, while decreasing the number of measurements in each interval, by doubling the timing  between successive measurements $\tau$. Again, the standard deviation $\sigma(\tau)$ would depend on $\tau$ only for $\tau>\tau_{eq}$. This is shown in Fig.~\ref{sigma_l48}, where we plotted $\sigma(\tau)$ for $L=48$, for the
correlation function $G(0,L/2)$. The equilibration time can be
estimated to be $\tau_{eq}\sim 2^{14}$. The equilibration times
estimated in this way for other sizes are summarized in the Table
\ref{Ltau}.

\begin{table}
\caption{Equilibration time}
\begin{tabular}{|c||c|c|c|c|}
\hline
$L$ & 8 & 16 & 32 & 64\\
\hline
$\tau$ & $2^4=16$ & $2^7=128$ & $2^{11}=2048$ & $\ge 2^{15}=32768$\\
\hline
\end{tabular}
\label{Ltau}
\end{table}



\begin{thebibliography}{99}
\bibitem{fn-topomoves}
Any two colour configurations connected by a sequence of such loop
moves contribute to the overlap with the same sign.  It may however be
the case that some topological sectors of the
four-colouring model cannot be connected by such loop moves. This
would, if anything, suppress the overlap for larger transition graph
elements.
\end{thebibliography}
\end{document}